\documentclass{moriond}
\usepackage{braket}

\usepackage{multirow}
\usepackage{placeins}

\def\be{\begin{equation}}
\def\ee{\end{equation}}
\def\bea{\begin{eqnarray}}
\def\eea{\end{eqnarray}}

\newcommand{\Photo}{}

\begin{document}
\vspace*{4cm}
\title{Muon g-2 in SUSY scenarios with and without stable neutralinos}

\author{ Rafa\l{} Mase\l{}ek }

\address{
Institute of Theoretical Physics, Faculty of Physics,\\
University of Warsaw, ul.~Pasteura 5, PL 02-093 Warsaw, Poland
}

\maketitle\abstracts{
We investigate a possibility to explain the discrepancy between the Standard Model predictions and the observed value of the anomalous magnetic moment of the muon within unorthodox SUSY scenarios in which neutralinos are unstable. We start by reviewing the muon g-2 calculations in the MSSM and confront it with the most up-to-date experimental constraints. We find out that the next generation of direct detection DM experiments combined with the current LHC results will ultimately test 
the parameter region that can explain the $(g-2)_\mu$ anomaly in the MSSM.
Next, we study R-parity violating and gauge-mediated SUSY-breaking scenarios with unstable neutralinos. 
These models do not provide a viable DM candidate, which allows them to evade the DM constraints.
We find that in RPV and GMSB with slepton NLSP the LHC constraints are weaker, and a large region of parameter space can explain the observed anomaly and evade experimental limits.
}

\section{Introduction}

One of the most interesting anomalies observed in the recent years is a $4\sigma$ discrepancy between the anomalous magnetic moment of the muon
calculated theoretically within the \textit{Standard Model (SM)} and
the value measured experimentally
at the Brookhaven National Laboratory \cite{Muong-2:2006rrc}  and confirmed in 2021
at the Fermilab \cite{Muong-2:2021ojo}. Taking the average of the two measurements and comparing it to the SM prediction taken from the
white paper \cite{Aoyama:2020ynm}
\footnote{There are, however, new lattice-based results differing from the previous theoretical calcullations \cite{Borsanyi:2020mff}  for which the discrepancy with the experiment is much smaller.}
, one obtains:
\begin{equation}\label{eq:amu}
\Delta a_\mu = a_\mu ^{BNL+FNAL} - a_\mu^{SM} = (25.1 \pm 5.9)\times 10^{-10},
\end{equation}
where $a_\mu = (g_\mu-2)/2$.

The anomaly can be explained by the \textit{Beyond the Standard Model (BSM)} Physics scenarios, e.g. models with leptoquarks, axion-like particles, vector-like leptons or \textit{Dark Matter (DM)}. One of the other explanations is \textit{Supersymmetry (SUSY)}, which already in its 
simplest form, the \textit{Minimal Supersymmetric Standard Model (MSSM)} \cite{Haber:1984rc},
not only solves the anomaly, but also provides cadidate for DM particle, coupling unification and radiative breaking of the electroweak symmetry.

Within the MSSM, the anomaly can be explained at the one loop level using six parameters: $\tan{\beta}$, $M_1$, $M_2$, $\mu$, $\tilde m_{l_L}$, 
$\tilde m_{l_R}$, where $\tan{\beta}=\braket{H^0_u}/\braket{H^0_d}$ is the ratio of vacuum expectation values of the two Higgs doublets, and other parameters are soft SUSY breaking masses of Bino, Wino, Higgsino, left-handed and right-handed slepton, respectively. In this study we postulate universal slepton masses in order to reduce the number of free parameters and to avoid strong constraint from the flavour violating processes, e.g. $\mu\to e \gamma$.

\begin{figure}[ht!]
\center
\includegraphics[width=\textwidth]{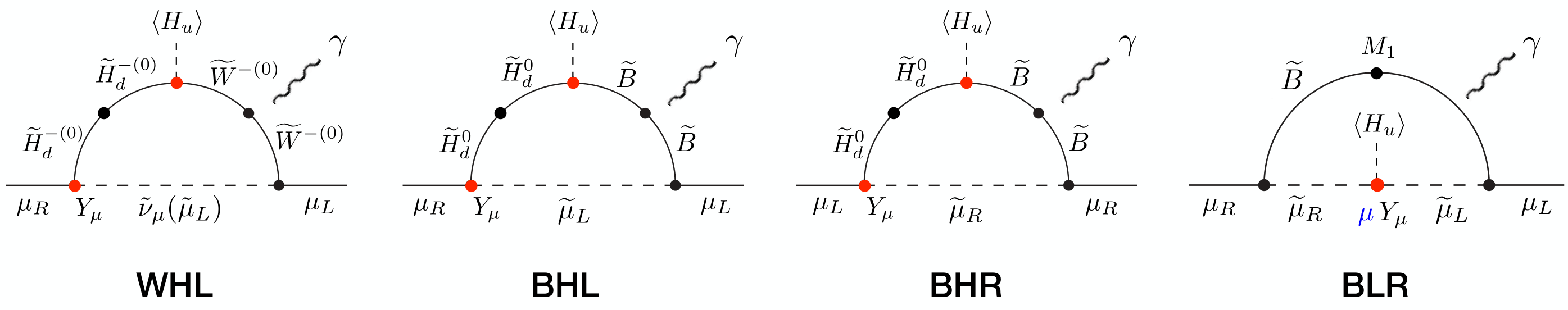}
\label{fig:diagrams}
\caption{One-loop diagrams in SUSY contributing to $a_\mu$. Red dots represent interactions responsible for the $\tan{\beta}$ enhancement.}
\end{figure}

At the one loop level, we can approximate the SUSY contribution to the anomalous magnetic moment of a muon by:
\begin{equation}\label{eq:contr}
a_\mu^{SUSY} \approx 
a_\mu^{WHL} + 
a_\mu^{BHL} + 
a_\mu^{BHR} + 
a_\mu^{BLR}, 
\end{equation}
where each term corresponds to the appropriate loop diagram depicted in Fig. \ref{fig:diagrams}. The labels represent SUSY particles entering the loop, e.g. WHL stands for Wino and Higgsino and left-handed slepton. When masses of the three relevant particles are of order of $\mathcal{O}(100~\mathrm{GeV})$ and other mass parameters are large, then contribution from the appropriate loop diagram dominates other terms in Eq. \ref{eq:contr}. 

\vskip -0.5em
Numerous studies targeting the SUSY contribution to $a_\mu$ have already been conducted \cite{Martin:2001st}, but most of them assume one of the neutralinos to be the lightest SUSY particle (LSP) and stable due to the R-parity conservation.
Such scenarios identify the LSP as a DM candidate and suffer from tight constraints coming from measurements of the spin-independent 
DM-nucleon scattering cross-section.
{In this study we aim at performing an analysis of SUSY scenarios in which neutralino is unstable, establishing viable SUSY $(g-2)\mu$ parameter region, and confrotning it with current experimental constraints.
}

\section{Analysis}

Results of the full study are described in the main article \cite{fullpaper}, where we have introduced eight two-dimensional parameter planes, two per each term in Eq. \ref{eq:contr}. In this paper we discuss only 
$WHL_\mu$ plane, where $M_2$ and $\mu$ are parameters of interest; 
$\tilde m_{l_L} = \rm{min}(M_2,\mu) + 20~\rm{GeV}$, $M_1=\tilde m_{l_R}=10~\rm{TeV}$ and $\tan{\beta}=50$.

We begin by reviewing the viable $(g-2)_\mu$ parameter region in the MSSM with neutralino LSP and checking for constraints from the LHC analyses and direct DM detection experiments. Next, we break the R-parity by introducing a baryon number violating $UDD$ operator. Breaking of the symmetry causes the LSP to decay to 3 (anti)quarks, and by the price of loosing a DM candidate, allows to evade all DM constraints. Finally, we investigate a \textit{gauge-mediated SUSY-breaking (GMSB)} scenario, in which the LSP is not neutralino, but very light (1 eV) gravitino. We take the 
\textit{next to the lightest SUSY particle (NLSP)} to be slepton/stau/sneutrino, because for neutralino NLSP the interesting parameter region is already totally excluded \cite{fullpaper}.

In order to perform the analysis, we generate over 900 SUSY mass points per plane using SUSY-HIT and SDECAY, and for each of them we calculate the SUSY contribution to the $a_\mu$ using GM2Calc. In order to recast the LHC analyses we use CheckMATE2. For the MSSM we calculate the neutralino relic density and spin-independent 
DM-nucleon scattering cross-section using MicrOmegas5.

\section{Results}

Figure 2 presents results of the analysis for the $\rm{WHL}_\mu$ plane and three SUSY scenarios: MSSM (left), R-parity violation (RPV) through $UDD$ operator (center), and GMSB (right) with neutralino NLSP. Each of the plots contains a green and yellow bands corresponding to $1\sigma$ and $2\sigma$ agreement between the value of the anomaly in Eq. \ref{eq:amu} and SUSY contribution to $a_\mu$. 
The beige region in the lower left corner of the plot corresponds to parameters for which the SUSY contribution to $a_\mu$ exceeds the global average of experimental measurements.
The hatched contour corresponds to DM overproduction. It is present only for the MSSM case, because in the RPV scenario there is no DM candidate, and in the GMSB the abundance of the gravitino is tiny. The gray contour shadows parameter region for which the predicted spin-independent neutralino-nucleon scattering cross-section is larger than current limit from the Xenon1T experiment. In addition, contours representing expected constraints from the next-generation DM direct detection experiments are shown, heavily constraining the MSSM plane. Other contours correpond to LHC contraints, which are listed in Tab. \ref{tb:analyses}.

\begin{table}[ht!]
\begin{center}
\caption{\label{tb:analyses}
The experimental analyses and simplified models used in 
the inspection of the MSSM, RPV and GMSB scenarios.
The colour indicates to the
region where the simplified model limit
provides 95\% CL exclusion in Fig. 2. The number of considered analyses was much larger, only the relevant ones are included.
}

\begin{tabular}{ |l|l|c|c|c|c| } 
\hline
Scenario & Analysis & $E/{\rm TeV}$ & ${\cal L}/{\rm fb}^{-1}$ & Colour \\
\hline
\multirow{3}{2em}{MSSM} &
CMS $\ell^+ \ell^-$ \cite{CMS:2020bfa} & 13 & 137 & Red \\ 
\cline{2-5}
& ATLAS soft-$\ell$ \cite{ATLAS:2019lng}  & 13 &139 & Blue \\ 
\cline{2-5}
& ATLAS DT \cite{ATLAS:2022rme} & 13 & 136~\, & Orange \\
\hline
\multirow{3}{2em}{RPV} 
& ATLAS multijet+$\ell$  \cite{ATLAS:2021fbt} & 13 & 139 & Red  \\
\cline{2-5}
& CMS multilepton  \cite{CMS:2017moi} & 13 & 35.9 & Blue  \\
\cline{2-5}
& ATLAS jets+$met$  \cite{ATLAS-CONF-2019-040} & 13 & 139 & Green  \\
\hline
\multirow{2}{2em}{GMSB} 
& CMS multilepton  \cite{CMS:2017moi} & 13 & 35.9 & Blue  \\
\cline{2-5}
& CMS soft $\ell^+ \ell^-$  \cite{CMS:2018kag} & 13 & 35.9 & Orange  \\
\hline

\end{tabular}
\end{center}
\end{table}
One can see that current experimental limits rule out a possiblity to explain the observed anomaly within the MSSM, except for two small regions around 
$(M_2=400~\mathrm{GeV},\mu=700~\mathrm{GeV})$ and $(M_2=1000~\mathrm{GeV},\mu=300~\mathrm{GeV})$. However, both of these regions will be completely tested by the next generation of the direct DM detection experiments. The situtation is much more interesting for the R-parity violating scenario. In this theory there is no DM candidate, hence all DM limits are gone, but some of the LHC analyses are sensitive, however, the parameter regions where one of the mass parameters is larger than 1 TeV, are uncostrained and still able to explain the measured value of the anomalous magnetic moment of the muon. Similarly to the RPV case, the gauge-mediated SUSY-breaking scenario is also free from DM constraints. In this model, parameter region with $M_2 < 450 ~ \mathrm{GeV}$ is excluded for all interesting values of the $\mu$ parameter, because of the CMS soft lepton search \cite{CMS:2018kag}. For larger values of the Wino mass parameter, GMSB SUSY scenario with stau NLSP can easily explain the latest Fermilab result.

\begin{figure}[ht!]\label{fig:results}
\centering
\includegraphics[width=0.31\textwidth]{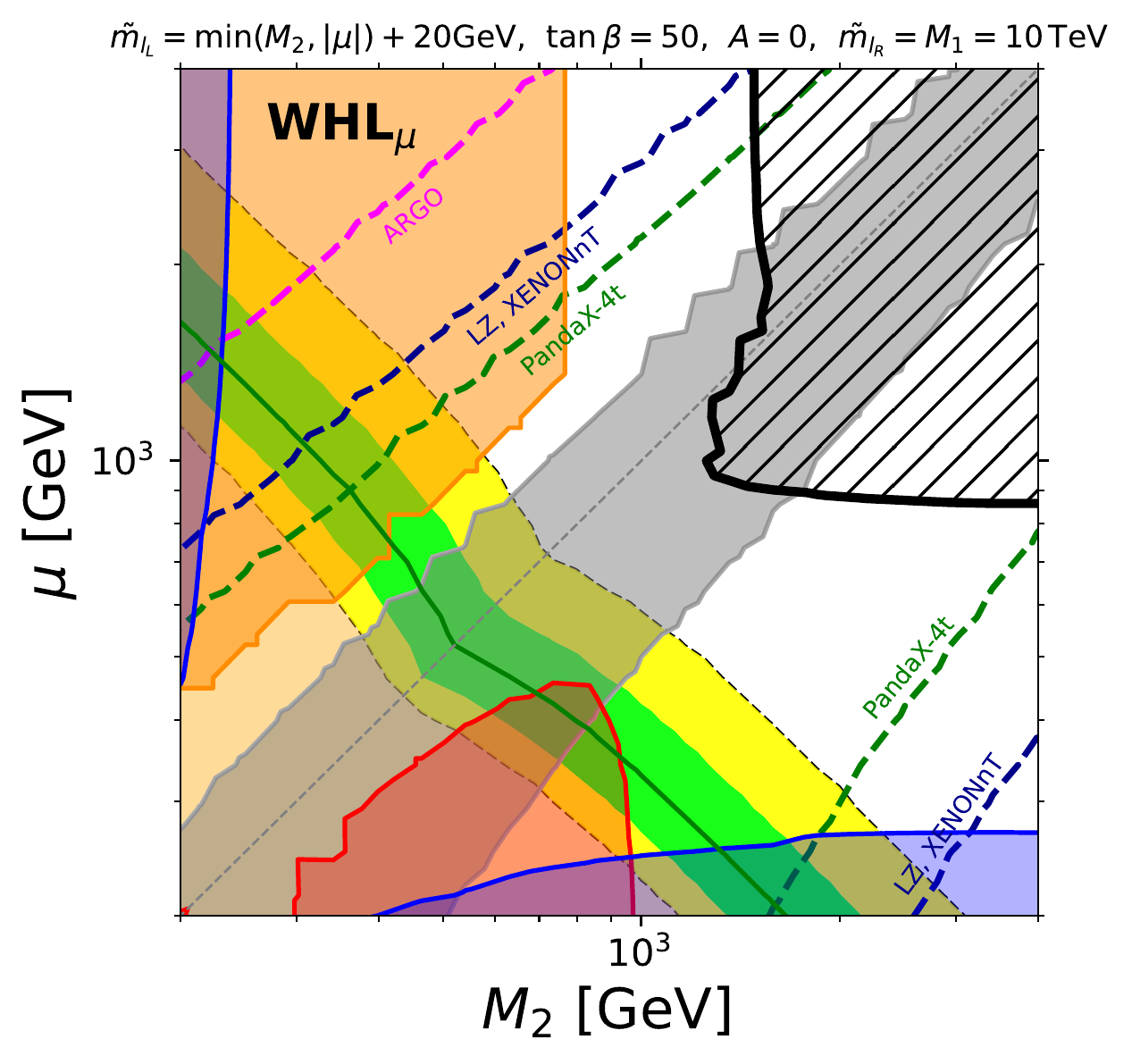}
\includegraphics[width=0.31\textwidth]{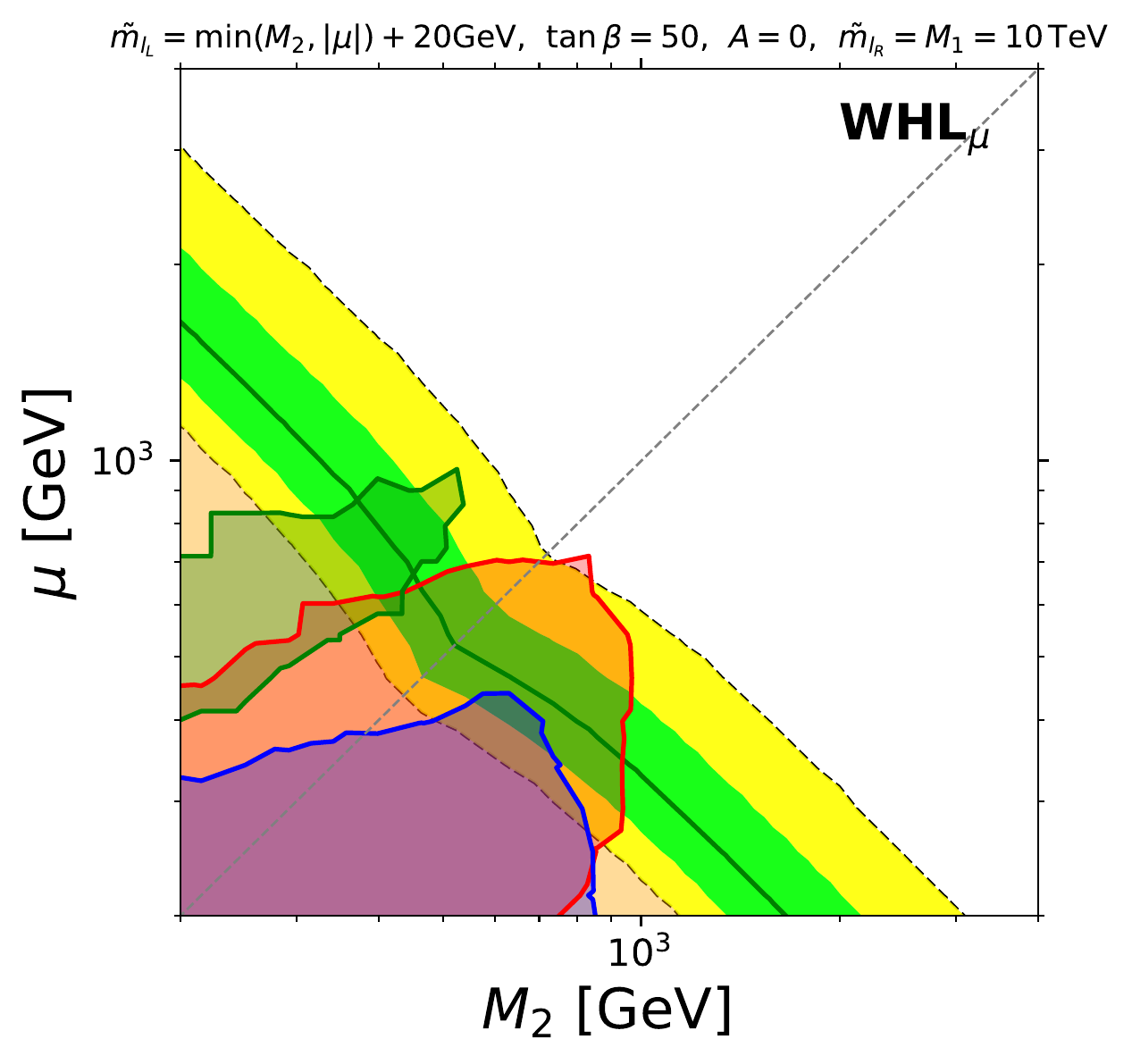}
\includegraphics[width=0.31\textwidth]{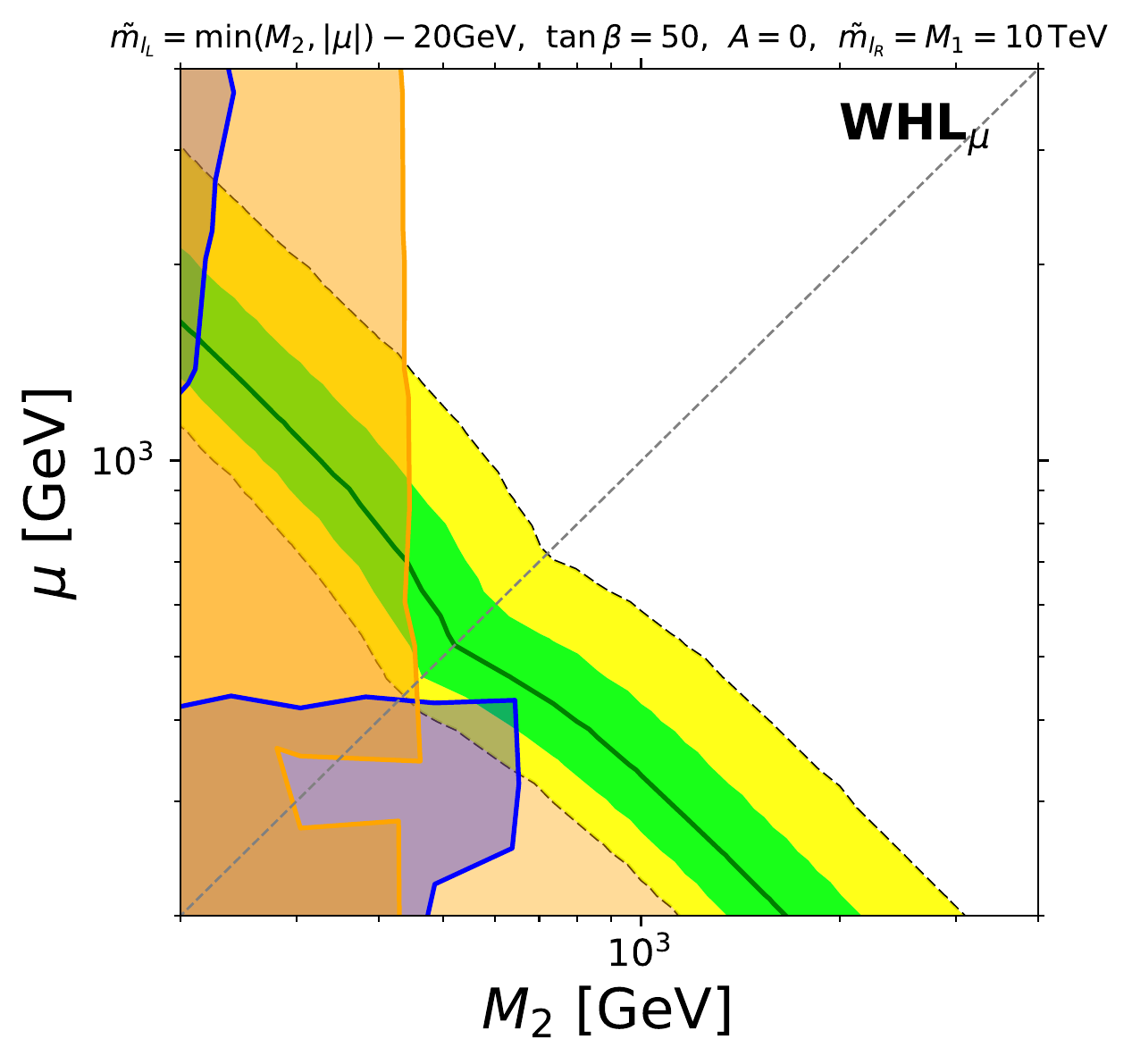}
\caption{ Results for the MSSM (left), R-parity violating (center), and GMSB (right) SUSY scenarios. Green and yellow bands correspond to $1\sigma$ and $2\sigma$ agreement with the value of the anomaly in Eq. \ref{eq:amu}. Shaded and hatched regions correspond to parameters excluded by current experimental results, see description in the text for details.
}
\end{figure}
\vskip -0.5em

\section{Conclusions}
In this article we discuss a part of the result presented in the main paper \cite{fullpaper}, aiming at providing an explanation of the observed value of the anomalous magnetic moment of the muon in
the MSSM and
SUSY scenarios without stable neutralinos. We concentrate on a plane, where we vary Wino and Higgsino mass parameters, assuming a 20 GeV mass degeneracy between the left-handed slepton and the lightest out of $M_2$ and $\mu$. While the most popular scenario, which is the MSSM with nautralino LSP being a natural DM candidate, is almost completely excluded by current experimental constraints, RPV and GMSB models with unstable neutralinos can explain the observed anomaly and evade experimental limits. The main reason is that these models lack the DM candidate, hence DM constraints do not apply. The other reason is that for the SUSY scenarios with unstable neutralino the region, which is interesting from the point of view of muon g-2 anomaly, is less constrained by the LHC analyses. We hope that our work will attract more attention to unorthodox SUSY scenarios with unstable neutralinos, and encourage experimental collaborations to improve collider limits. 

\section*{Acknowledgments}
The work of R.M.\
is partially supported by the National Science Centre, Poland,
under research grant 2017/26/E/ST2/00135
and the Beethoven grant DEC-2016/23/G/ST2/04301.

\section*{References}
\small
\bibliography{ref}

%
%
%
%
%

\end{document}